\documentclass[sigconf,screen]{aamas}

\usepackage{balance} %
\usepackage[nameinlink,capitalise]{cleveref} %

\usepackage{makecell}
\usepackage{multirow}

\setcopyright{ifaamas}
\acmConference[AAMAS '24]{Proc.\@ of the 23rd International Conference
on Autonomous Agents and Multiagent Systems (AAMAS 2024)}{May 6 -- 10, 2024}
{Auckland, New Zealand}{N.~Alechina, V.~Dignum, M.~Dastani, J.S.~Sichman (eds.)}
\copyrightyear{2024}
\acmYear{2024}
\acmDOI{}
\acmPrice{}
\acmISBN{}

\newcommand{\our}{{HiMAP}}

\acmSubmissionID{543}

\def \papertitle{\our{}: Learning Heuristics-Informed Policies for Large-Scale Multi-Agent Pathfinding}

\title[\papertitle]{\papertitle}
\subtitle{Extended Abstract}

\author{Huijie Tang$^*$}
\thanks{$^*$Equal contributions.}
\affiliation{
  \institution{KAIST}
  \city{Daejeon}
  \country{South Korea}}
\email{raylan.tang@kaist.ac.kr}

\author{Federico Berto$^*$}
\affiliation{
  \institution{KAIST}
  \city{Daejeon}
  \country{South Korea}}
\email{fberto@kaist.ac.kr}

\author{Zihan Ma}
\affiliation{
  \institution{KAIST}
  \city{Daejeon}
  \country{South Korea}}
\email{zihanma@kaist.ac.kr}

\author{Chuanbo Hua}
\affiliation{
  \institution{KAIST}
  \city{Daejeon}
  \country{South Korea}}
\email{cbhua@kaist.ac.kr}

\author{Kyuree Ahn}
\affiliation{
  \institution{Omelet}
  \city{Daejeon}
  \country{South Korea}}
\email{ahnkjuree@gmail.com}

\author{Jinkyoo Park}
\affiliation{
  \institution{KAIST, Omelet}
  \city{Daejeon}
  \country{South Korea}}
\email{jinkyoo.park@kaist.ac.kr}

\begin{abstract}
Large-scale multi-agent pathfinding (MAPF) presents significant challenges in several areas, such as autonomous vehicle management in smart cities and warehouse robot control. As systems grow in complexity, efficient and collision-free coordination becomes paramount. Traditional algorithms often fall short in scalability, especially in intricate scenarios. Reinforcement Learning (RL) has shown potential to address the intricacies of MAPF; however, it has also been shown to struggle with scalability, demanding intricate implementation, lengthy training, and often exhibiting unstable convergence, limiting its practical application. In this paper, we introduce Heuristics-Informed Multi-Agent Pathfinding (\our{}), a novel scalable approach that employs imitation learning with heuristic guidance in a decentralized manner. We train on small-scale instances using a heuristic policy as a teacher that maps each single agent observation information to an action probability distribution. During pathfinding, we adopt several strategies to improve performance such as \textit{Preventing Re-Visit}, \textit{Treating Completed Agents as Obstacles} and \textit{Softmax Temperature Adaptation}. With a simple training scheme and implementation, \our{} demonstrates competitive results in terms of success rate and scalability in the field of imitation-learning-only MAPF, showing the potential of imitation-learning-only MAPF
equipped with inference techniques.
\end{abstract}


\newcommand{\BibTeX}{\rm B\kern-.05em{\sc i\kern-.025em b}\kern-.08em\TeX}

\makeatletter
\gdef\@copyrightpermission{
	\begin{minipage}{0.3\columnwidth}
		\href{https://creativecommons.org/licenses/by/4.0/}{\includegraphics[width=0.90\textwidth]{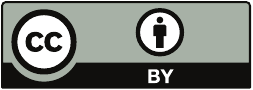}}
	\end{minipage}\hfill
	\begin{minipage}{0.7\columnwidth}
		\href{https://creativecommons.org/licenses/by/4.0/}{This work is licensed under a Creative Commons Attribution International 4.0 License.}
	\end{minipage}
	\vspace{5pt}
}
\makeatother

\begin{document}

\hypersetup{
    linkcolor=magenta,
    citecolor=green!80!black,
    urlcolor=cyan,
}

\pagestyle{fancy}
\fancyhead{}

\maketitle 



\section{Introduction}

Multi-Agent Pathfinding (MAPF) entails finding collision-free paths for a group of agents while minimizing makespan or flowtime \citep{stern2019multi, ahn2023schedulingmapf}. MAPF is a domain with several practical applications in our increasingly automated world, including warehouse robotics \cite{wurman2008coordinating}, aviation \cite{morris2016planning}, and digital gaming \cite{ma2017feasibility} scenarios. %

Centralized strategies have offered solutions by reducing MAPF to other familiar problems, e.g., ILP \cite{yu2013planning} and SAT \cite{surynek2016efficient}, or employing search-based algorithms, e.g., Conflict-Based Search (CBS) \cite{sharon2015conflict} and its improved versions ECBS \cite{barer2014suboptimal} and EECBS \cite{li2021eecbs}. Yet, scalability remains hard to obtain, especially as the number of agents increases. %

Reinforcement Learning (RL) approaches \citep{ma2021learning, ma2021distributed, lin2023sacha} offer another way to solve the MAPF problem by treating it as a sequential decision-making problem rather than a centralized planning problem. However, these approaches often involve intricate implementation and lengthy training and may exhibit unstable convergence unless imitation learning is also used to stabilize convergence \citep{sartoretti2019primal, damani2021primal, chen2022multiagent, wang2023scrimp}.

\textbf{\textit{Contributions}}. In this paper, we propose Heuristics-Informed Multi-Agent Pathfinding (\our{}), a novel method that relies only on imitation learning (IL) through which the neural network model learns to act from heuristics expert solutions. Our contributions include: 1) We formulate the MAPF problem as an imitation learning problem where the model learns to choose actions by imitating expert decisions. 2) We propose a simple yet effective training scheme based on imitation learning of small-scale solutions from heuristics solvers. 3) We introduce several inference techniques to improve the performance. 4) We demonstrate the potential of imitation-learning-only MAPF equipped with inference techniques.

\section{\our{}}
\our{} trains with the help of heuristics guidance in IL manner. During pathfinding, \our{} first gathers observations from all agents, where each observation comprises four distinct channels. Subsequently, these observations are fed into the neural network. Leveraging four distinct inference techniques, the neural network produces action for all agents. As a result, the locations of all agents are updated, facilitating a transition to the subsequent state.

\textbf{\textit{Network Architecture and Observation Space}}. We adopt the same neural network as PRIMAL \cite{sartoretti2019primal}, while eliminating the Value and Blocking output head. Similar to PRIMAL, we use a partially observable grid world as the environment. Each agent $i$ at location $v_i^t$ has a four-channel observation $o_i^t$ in a $9 \times 9$ Field of View (FOV) centered around itself, plus a $3 \times 1$ vector pointing from its own position to its goal. At each time step $t$, agents act in the grid world either to move to an adjacent vertex or stay at the current vertex.

\textbf{\textit{Heuristic Guidance}}. We train the model with the help of heuristic expert guidance in an imitation learning (IL) manner. We use EECBS \cite{li2021eecbs} to generate a series of expert collision-free paths on randomly generated environments. In the generated heuristic expert paths, each location of a single agent $i$ at time $t$, i.e., $v_i^t$, can be viewed as a decision point, and the decision ${a^t_i}^*$ according to the expert paths is either up, down, left, right, or stay still. \our{} is trained by minimizing the MSE loss between the model output $\hat{a}_i^t$ given the input observation $o_i^t$, and the expert action decision ${a^t_i}^*$.

\textbf{\textit{Inference Techniques}}. 1) \textit{Preventing Re-Visit}: 
During the pathfinding, we find that sometimes agents cannot plan well due to their limited FOV. This will leave some agents oscillating around dead ends and never reaching their goals. Hence, we introduce a history recording scheme that records the history of visited locations and prevents the agents from repetitively re-visiting visited locations during the last few time steps. %

2) \textit{Softmax Temperature Adaptation}: We found that planning with only greedy inference does not perform well. Therefore, we adjust the softmax temperature of the final layer of the model such that it can increase or decrease the amount of exploration. During training, the temperature is set to $\tau = 1$. During pathfinding, we can adjust the value of $\tau$ to induce more or less exploration of the map, which can be beneficial for the agent to get unstuck.

3) \textit{Treating Completed Agents as Obstacles}: In this work, we adopt the \textit{stay at target} setting, as defined in \citet{stern2019multi}. Thus, agent overlapping with its goal can be effectively treated as an obstacle from the model's perspective. Therefore, we propose changing the map representation for completed agents into obstacles to improve the model performance effectively. We observe that results are improved compared with the situation without this modification, especially with many agents.

4) \textit{Conflict-Free Planning}: We consider four types of conflicts: \textit{swapping conflict} \cite{stern2019multi}, \textit{vertex conflict} \cite{stern2019multi}, conflict with static obstacles, and out-of-bound conflict. During action sampling, the sampled action may guide agents to take invalid actions, leading to one of the conflicts. We implement a two-stage technique to correct invalid actions to enable conflict-free planning. Before sampling, we identify the actions that may result in agents colliding with obstacles or going out of bounds and mask these actions by setting their probability to $0$ so they do not influence decision-making. After sampling, the sampled actions may still result in \textit{swapping conflict} or \textit{vertex conflict}. Similar to \citet{ma2021distributed}, we employ simple rules to avoid these two conflicts. If these two conflicts happen, the related agents' states are recursively recovered until no conflicts exist.

\section{Experiments}
\textit{Training Data Generation}. We generate 100 random square maps with sizes $40 \times 40$ and obstacle density 0.3, and another 100 maps with sizes $80 \times 80$ and the same obstacle density. For each map, we create five \textit{scenario files} \cite{stern2019multi}, each containing 4, 8, 16, 32, or 64 start-goal location pairs, resulting in a total of 1000 \textit{scenario files}. We solve all 1000 Multi-Agent Path Finding (MAPF) problems using EECBS with a suboptimality factor of 1.2. The training dataset consists of pairs $(o_i^t, {a_i^t}^*)$, generated from expert solutions. Generating all expert paths takes less than a minute on an \textsc{Intel Core i7-6850K}. Other hyperparameters include a learning rate starting at $5 \times 10^{-5}$, decreasing by 80\% every eight epochs, and a total of 35 training epochs. Training takes approximately 100 minutes on a single \textsc{NVIDIA TITAN X} GPU with 12GB of VRAM. Reproducible source code is available on GitHub\footnote{\href{https://github.com/kaist-silab/HiMAP}{https://github.com/kaist-silab/HiMAP}}.

 We follow \citet{chung2023learning} and define the success rate as the average percentage of agents reaching the destination without collisions out of the total number of agents within the maximum allowed time steps. The maximum allowed time step during testing is 256 for the $40 \times 40$ map and 386 for the $80 \times 80$ map. For pathfinding, we use history size $H = 5$, i.e., the agent is banned from re-visiting the location it has visited in the last five time steps, and softmax temperature $\tau=2$.

\begin{figure}
    \centering
    \includegraphics[width=\linewidth]{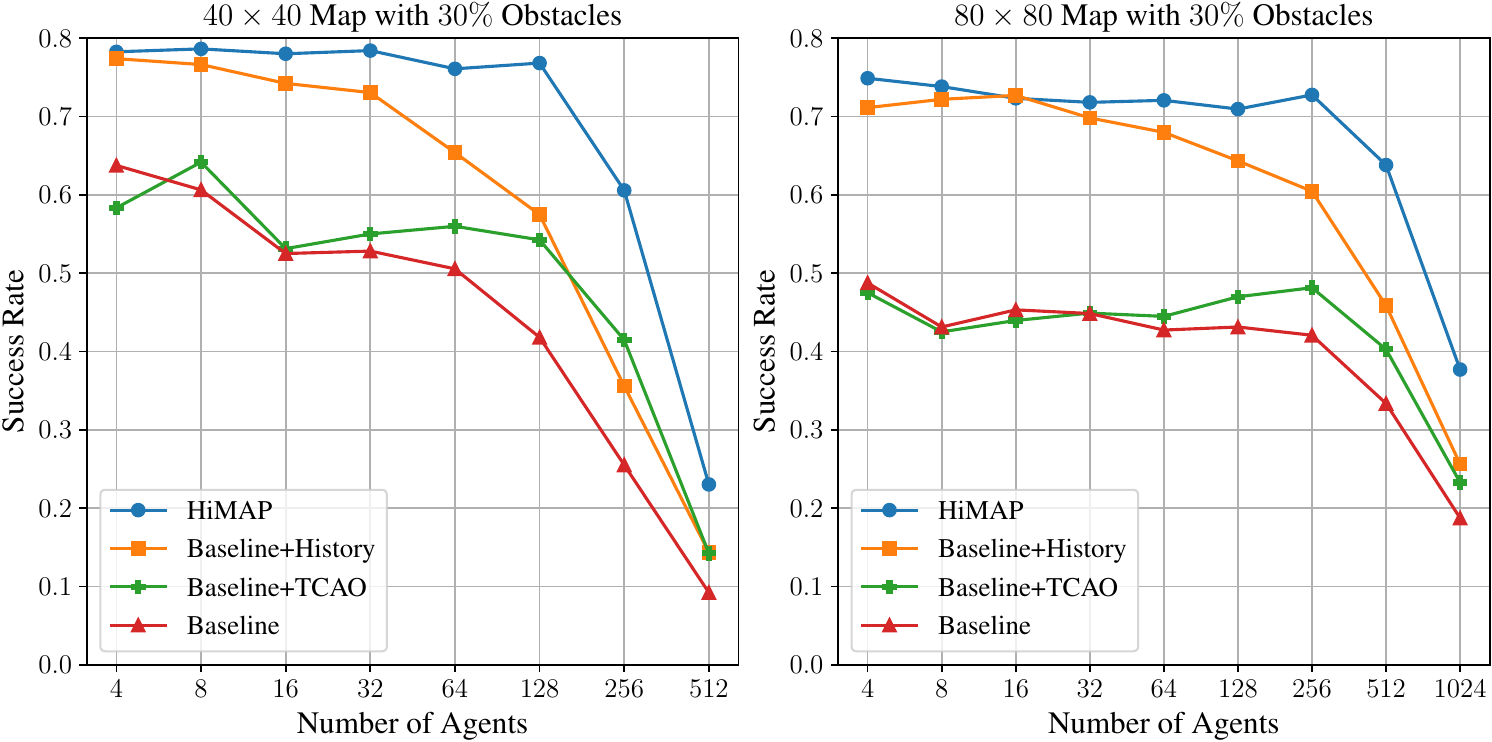}
    \caption{Success rate of \our{} and ablated models. Baseline: \our{} without preventing re-visit (No History) and treating completed agents as obstacles (No TCAO). }
    \label{fig:main_res}
\vspace{-1.3em}
\end{figure}

The success rate is shown in \cref{fig:main_res}, demonstrating the power of \our{} with pure IL equipped with inference techniques. We also show the success rate of ablated models to verify the effectiveness of proposed inference techniques. We believe that the results can be further improved by using a larger training dataset, which will be left for future work.

\section{Discussion and Conclusion}
\paragraph{Discussion}
To the best of our knowledge, \citet{abreu2022efficient} is the only method that employs pure IL to train neural network solvers for MAPF, using a dataset based on single-agent shortest paths \cite{ren2021mapfast}. However, these paths do not guarantee collision-free paths for agents or allow agents to observe each other's behavior. In contrast, equipped with inference techniques, \our{} uses collision-free paths generated by EECBS, which better represents the MAPF solution in dynamic settings.

\paragraph{Conclusion}
We proposed \our{}, an IL approach that employs heuristics expert paths to solve MAPF problems, whose implementation and training are easier and less costly compared with RL-based solvers. The training dataset only contains paths on relatively small-scale MAPF problems, making it easy to collect expert paths with existing heuristics solvers. We further introduced several inference techniques to boost performance during evaluation. We demonstrate \our{}'s success rate in different environments with high obstacle density, showcasing the potential of imitation-learning-only MAPF equipped with inference techniques. Although, at the current stage, \our{} cannot outperform state-of-the-art learning-based techniques, we believe imitation-based approaches offer an interesting avenue for future works. In particular, policies hybridizing domain knowledge and possibly other inference techniques as in \citet{gao2023rde} could further improve IL-based MAPF. Finally, learning to communicate as in \citet{ma2021learning, wang2023scrimp} is a potential area that could drastically improve \our{}'s performance.

\bibliographystyle{ACM-Reference-Format} 
\bibliography{bibliography}

\end{document}